\documentclass[aps,twocolumn,preprintnumbers,amsmath,amssymb,nofootinbib,superscriptaddress,notitlepage]{revtex4-1}

\usepackage{epsfig}
\usepackage{txfonts}
\usepackage{soul}
\usepackage{color}
\usepackage{bm}
\usepackage{braket}

\newcommand{\mpv}[1]{#1}
 
\begin{document}

\title{Heavy-hadron molecules from light-meson-exchange saturation}

\author{Fang-Zheng Peng}
\affiliation{School of Physics,  Beihang University, Beijing 100191, China}

\author{Ming-Zhu Liu}
\affiliation{School of Space and Environment, Beihang University, Beijing 100191, China}
\affiliation{School of Physics,  Beihang University, Beijing 100191, China}

\author{Mario {S\'anchez S\'anchez}}
\affiliation{Centre d'\'Etudes Nucl\'eaires, CNRS/IN2P3, Universit\'e de Bordeaux, 33175 Gradignan, France}

\author{Manuel {Pavon} Valderrama}\email{mpavon@buaa.edu.cn}
\affiliation{School of Physics,  Beihang University, Beijing 100191, China}
\affiliation{International Research Center for Nuclei and Particles
  in the Cosmos \&
  Beijing Key Laboratory of Advanced Nuclear Materials and Physics,
  Beihang University, Beijing 100191, China}

\date{\today}


\begin{abstract} 
  \rule{0ex}{3ex}
  In the effective field theory framework the interaction
  between two heavy hadrons can be decomposed
  into a long- and a short-range piece.
  The long-range piece corresponds to the one-pion-exchange potential
  and is relatively well-known.
  The short-range piece is given by a series of contact-range interactions
  with unknown couplings, which substitute the less well-known
  short-range dynamics.
  While the general structure of the short-range potential
  between heavy hadrons
  is heavily constrained from heavy-quark symmetry, the couplings
  are still free parameters.
  Here we argue that the relative strength and the sign of these couplings
  can be estimated from the hypothesis that they are saturated
  by the exchange of light mesons, in particular the vector mesons $\rho$
  and $\omega$, i.e. from resonance saturation.
  However, we propose a novel saturation procedure
  that effectively removes form-factor artifacts.
  From this we can determine in which spin and isospin configurations
  the low{-}energy constants are most attractive
  for specific two{-}heavy-hadron systems.
  In general the molecular states with lower {isospins} and higher spins
  will be more attractive and thus more probable candidates to form
  heavy-hadron molecules.
  This pattern is compatible with the interpretation of
  the $X(3872)$ and $P_c(4312/4440/4457)$ as molecular states,
  but it is not applicable to states
  with maximum isospin like the $Z_c(3900/4020)$.
\end{abstract}

\maketitle

Heavy-hadron molecules might very well be the most popular type of
exotic hadron~\cite{Voloshin:1976ap,DeRujula:1976qd,Guo:2017jvc}.
The probable reason is their conceptual simplicity,
which is only matched by the challenge 
{of making} concrete
predictions in the molecular picture.
Despite just being non-relativistic bound states of two heavy hadrons,
the theoretical toolbox behind hadronic molecules has grown
into a bewildering hodgepodge
which is often difficult to disentangle, to say the least.
This is in contrast with the much more coherent descriptions offered
by the quark model~\cite{Godfrey:1985xj,Capstick:1986bm} or the theory behind
quarkonium~\cite{Eichten:1978tg,Eichten:1979ms,Brambilla:1999xf,Brambilla:2004jw,Brambilla:2010cs}.

Yet the molecular picture has a few remarkable successes under its sleeves.
They include the prediction of
the $X(3872)$ by T\"ornqvist~\cite{Tornqvist:1993ng},
later detected by the Belle collaboration~\cite{Choi:2003ue},
and the prediction of three hidden{-}charm pentaquarks~\cite{Wu:2010jy,Wu:2010vk,Wu:2010rv,Xiao:2013yca,Wang:2011rga,Yang:2011wz,Karliner:2015ina}
($\Sigma_c \bar{D}$ and $\Sigma_c \bar{D}^*$ molecules),
which might very well correspond with the $P_c(4312)$, $P_c(4440)$ and
$P_c(4457)$ pentaquarks recently detected by the LHCb~\cite{Aaij:2019vzc}.
Regarding the $X(3872)$, the most compelling evidence that it is molecular
is not necessarily its closeness to the $D^* \bar{D}$
threshold~\cite{Tornqvist:2003na,Voloshin:2003nt,Braaten:2003he}
but its isospin{-}breaking decays~\cite{Choi:2011fc} which are
naturally reproduced in the molecular
picture~\cite{Gamermann:2009fv,Gamermann:2009uq}
(in the non-molecular case this feature
might~\cite{Swanson:2003tb} or might not be explainable~\cite{Hanhart:2011tn}
depending on the details of the model).
For the LHCb pentaquarks, though the molecular explanation is gaining
traction~\cite{Roca:2015dva,He:2015cea,Xiao:2015fia,Chen:2015loa,Chen:2015moa,Burns:2015dwa,Geng:2017hxc,Chen:2019bip,Chen:2019asm,Liu:2018zzu,Liu:2019tjn,Xiao:2019aya,Valderrama:2019chc,Burns:2019iih,Pan:2019skd,Du:2019pij},
there are a few competing hypotheses
about their nature~\cite{Eides:2019tgv,Wang:2019got,Cheng:2019obk}.

Despite the numerous candidates and the intense theoretical interest,
the qualitative and quantitative properties of the molecular
spectrum are poorly understood.
The present manuscript attempts to address this limitation
by proposing a potential pattern in the spectrum of
two{-}heavy-hadron bound states:
for configurations without maximum isospin, the states with higher
(light-quark) spin are expected to be lighter (i.e. more bound).
This is the opposite pattern as with compact hadrons,
for which mass usually increases with spin.
This pattern might explain why besides the $X(3872)$ no other
$D^{(*)}\bar{D}^{(*)}$ molecule has been observed,
as they should not be expected to be bound (with the exception of the
$2^{++}$ $D^* \bar{D}^*$ configuration~\cite{Valderrama:2012jv,Nieves:2012tt},
modulo other effects that could {unbind}
it~\cite{Cincioglu:2016fkm,Baru:2016iwj}).
If applied to the light sector, it also explains why in the two-nucleon system
the deuteron {binds} while the singlet state {does} not,
or why if the $d^*(2380)$~\cite{Adlarson:2011bh}
is a $\Delta \Delta$ bound state~\cite{Dyson:1964xwa}
{its spin} should be $J=3$.
It also states that if the $P_c(4440)$ and $P_c(4457)$ are 
$\Sigma_c \bar{D}^*$ bound states,
their expected quantum numbers are $\tfrac{3}{2}^-$ and
$\tfrac{1}{2}^-$, respectively.
This prediction, which agrees with a few theoretical analyses~\cite{Valderrama:2019chc,Liu:2019zvb,Du:2019pij},
will be put to the test by the eventual experimental determination of
the quantum numbers of the pentaquarks.

This pattern is deduced from matching a contact-range description of
the interaction between two heavy{ }hadrons with a phenomenological
description in terms of the potential generated
by the exchange of light{ }mesons.
That is, we are considering the saturation of the low-energy constants
by light-meson exchange (as in Refs. \cite{Ecker:1988te,Epelbaum:2001fm}).
We will illustrate this idea with the one-pion-exchange (OPE) potential,
which for two spin-$\tfrac{1}{2}$, isospin-$\tfrac{1}{2}$ hadrons reads
\begin{eqnarray}
   V(\vec{q}\,) &=&
  -\frac{g^2}{2 f^2}\,{\tau}\,\frac{\vec{\sigma}_1 \cdot \vec{q} \,
    \vec{\sigma}_2 \cdot \vec{q}}{q^2 + m_{\pi}^2}
  \nonumber \\
  &=& -\frac{g^2}{2 f^2}\,{\tau}\,\left[
    \frac{\frac{1}{3}\,\vec{\sigma}_1 \cdot \vec{\sigma}_2 q^2}{q^2 + m_{\pi}^2} +
    \frac{\vec{\sigma}_1 \cdot \vec{q} \, \vec{\sigma}_2 \cdot \vec{q} - \frac{1}{3}
    \vec{\sigma}_1 \cdot \vec{\sigma}_2 q^2}{q^2+m_{\pi}^2}
    \right] \, \nonumber,  \\\label{eq:OPE}
\end{eqnarray}
with $g$ the axial coupling, $f \sim 130\,{\rm MeV}$ {the pion decay constant},
$\vec{q}$ the exchanged momentum and $q = |\vec{q}\,|$,
$m_{\pi}$ the pion mass, {$\vec{\sigma}_{i}$ ($\bm{\tau}_i$) the Pauli matrices
for hadron $i=1,2$ in spin (isospin) space,} and {$\tau = \bm{\tau}_1 \cdot \bm{\tau}_2$}
an isospin factor.
In the second line the potential has been decomposed
into a spin-spin and a tensor piece.
We will ignore the tensor piece, as it requires {$\text{SD}$-}wave mixing.
We will consider the effect of OPE on the saturation of the couplings of
the lowest{-}order contact-range potential, which is purely S-wave.
\mpv{
  Finally we will ignore the practical and theoretical considerations
  derived from the fact that the pion is the lightest hadron
  (namely chiral symmetry):
  obviously under most settings we are not interested
  in saturation by pions, but in saturation
  by scalar- and vector-meson exchange.
  The choice of pions is merely intended as a simple example of
  the mechanics of saturation.
  }

\begin{figure}[ttt]
\begin{center}
  \includegraphics[width=9.2cm]{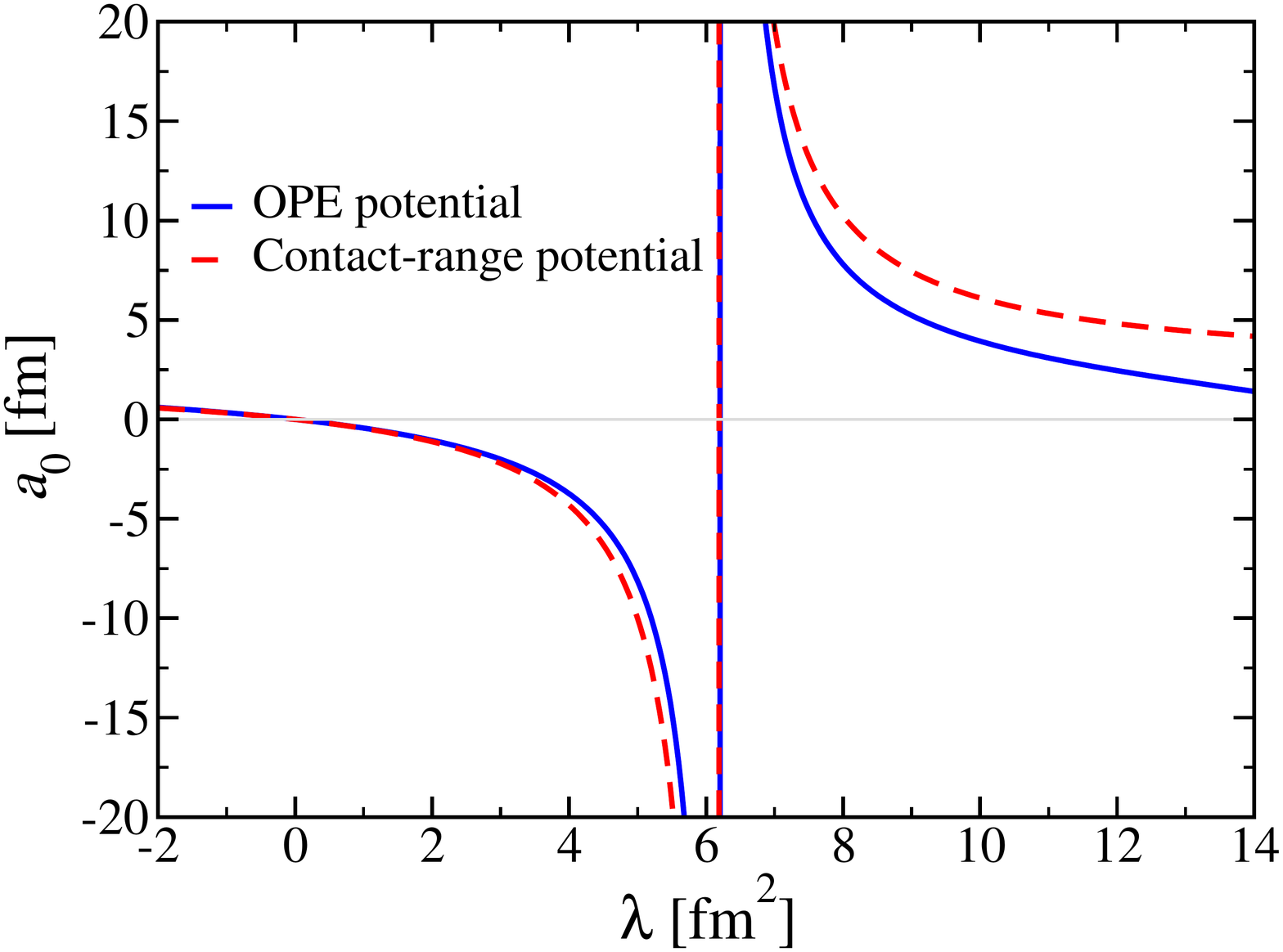}
\end{center}
\caption{
  Scattering length $a_0$ (in fm) for the S-wave {piece} of the OPE potential 
  {and} for the S-wave contact-range of Eq.~(\ref{eq:VC}) with
  the saturation conditions of Eq.~(\ref{eq:sat-OPE})
  and a regularization scale $\mu = 131\,{\rm MeV}$
  (close to the pion mass).
  We plot $a_0$ as a function of the strength of the potential 
  $\lambda = -\tau g^2 / 6 f^2 \vec{\sigma}_1 \cdot \vec{\sigma}_2$.
  The contact-range potential is regularized with a separable sharp{-}cutoff
  regulator in momentum space, i.e. $\braket{\vec{p}\,' | V_C | \vec{p}\,}
  = (C_0 + C_1\,\vec{\sigma}_1 \cdot \vec{\sigma}_2)\,
  \theta\left({{{\mu-p'}}}\right) \theta\left({{{\mu-p}}}\right)$.
}
\label{fig:saturation}
\end{figure}

The idea behind saturation is to map the \mpv{previous finite-range} potential
into an effective potential of the type
\begin{eqnarray}
  V_C(\vec{q}\,) = C_0(\mu) + C_1(\mu)\,\vec{\sigma}_1 \cdot \vec{\sigma}_2 \, ,
  \label{eq:VC}
\end{eqnarray}
which requires a regulator (not explicitly written here),
with {$\mu$ being} a regularization scale (i.e. a cutoff),
which we will choose around the mass of the exchanged
light meson ($\mu \sim m_{\pi}$ in this case)
for saturation to work.
If we expand the spin-spin piece of Eq. \eqref{eq:OPE} in powers of $q$,
\begin{eqnarray}
  V(\vec{q}\,) = - \frac{g^2}{6 f^2}\,\tau\,
  \vec{\sigma}_1 \cdot \vec{\sigma}_2
  \left[ \frac{q^2}{m_{\pi}^2} - \frac{q^4}{m_{\pi}^4} + \dots \right] \, ,
\end{eqnarray}
then, by matching this expansion with the effective potential $V_C$,
we will deduce that OPE should not saturate the couplings: 
\begin{eqnarray}
  C^{\rm OPE}_0(\mu \sim m_{\pi}) \sim 0\, , \quad
  C^{\rm OPE}_1(\mu \sim m_{\pi}) \sim 0 \, .
\end{eqnarray}
However this conclusion is premature.
If we rewrite the $q^2$-dependence as
\begin{eqnarray}
  \frac{q^2}{q^2 + m_{\pi}^2} = 1 - \frac{m_{\pi}^2}{q^2 + m_{\pi}^2} \, ,
\end{eqnarray}
then the first contribution in the right-hand side is actually a Dirac delta.
Owing to the finite size of the pions, this Dirac delta will acquire
a finite size $\sim 1/M$, with $M$ the physical cutoff of
the theory (probably a bit above $1\,{\rm GeV}$).
This does not necessarily coincide with the scale $\mu$
we use for the effective interaction.
In general saturation works best for $\mu \sim m$ with $m$ the mass of
the light meson, while for the exchange of a light meson to have
physical meaning we need $m < M$.
From this the saturation scale verifies $\mu < M$,
{implying that in practice} we can simply ignore contributions
with a range shorter than $1/\mu$ ($\sim 1/m$),
including the aforementioned delta.
Thus for saturation purposes we will simply make the substitution
\begin{eqnarray}
  \frac{q^2}{q^2 + m_{\pi}^2} \to - \frac{m_{\pi}^2}{q^2 + m_{\pi}^2} \, ,
  \label{eq:subs-pseudo}
\end{eqnarray}
in the exchange potential, \mpv{leading to
\begin{eqnarray}
 V(\vec{q}\,) &\to&
 \frac{g^2}{6 f^2}\,{\tau}\,
 \frac{\vec{\sigma}_1 \cdot \vec{\sigma}_2 \, m_{\pi}^2}
      {q^2 + m_{\pi}^2} + \dots \, ,
\end{eqnarray}
where the dots represent terms mixing S- and D-waves.
Matching at $q^2 = 0$, we obtain the saturated couplings:}
\begin{eqnarray}
  C^{\rm OPE}_0(\mu \sim m_{\pi}) \sim 0
  \,, \quad
  C^{\rm OPE}_1(\mu \sim m_{\pi}) \sim \frac{g^2}{6 f^2}\,\tau \, .
  \label{eq:sat-OPE}
\end{eqnarray}
Finally we can compare how well does the saturated contact-range interaction
versus the potential from which it is derived.
This is done in Fig. \ref{fig:saturation}, where we check that it works
relatively well for the scattering length $a_0$ as a function of
the strength of the potential 
{\mpv (see Appendix \ref{app:pionless} for supplementary details).}
Particularly saturation correctly reproduces the existence of a bound state,
which is signaled by a change of sign in \mpv{ $1/a_0$.}

\mpv{
  As previously noted,
  the saturation of low-energy couplings by OPE serves an illustrative purpose.
  Its practical value is limited: owing to chiral symmetry
  the pion mass is considerably lower than any other hadronic scales.
  In most practical settings, pion exchanges will be included explicitly
  as the finite-range potential, while the contact-range potential
  will be saturated by scalar- and vector-meson exchange.
  Pion saturation might be useful for the few hadronic molecules
  in which all the relevant momentum scales are lighter than the pion.
  With the exception of the deuteron {\mpv (or, more generally,
    few-nucleon systems)} and the $X(3872)$, which can be described
  in terms of a pionless EFT~\cite{vanKolck:1998bw,Chen:1999tn,Braaten:2003he,Fleming:2007rp},
  most hadronic molecules do not fall into this category~\cite{Valderrama:2012jv,Lu:2017dvm}
  (and even the few that fall might still benefit
  from a pionful treatment).
  Therefore the problem is to apply saturation to other light mesons,
  in particular the sigma, the rho and the omega.
}

\mpv{We will now explain the concrete application of saturation
  to heavy-hadron molecules.}
Instead of using the standard superfield formalism 
we will write
the interaction between two heavy hadrons in the light-quark
formalism described in Ref.~\cite{Valderrama:2019sid}
{\mpv(see Appendix \ref{app:H} for a more detailed explanation)}.
\mpv{This formalism merely amounts to notice that in the heavy-quark limit
  interactions among heavy hadrons do not depend on heavy-quark spin,
  which means that all spin dependence can be rewritten
  in terms of the spin degrees of freedom of
  the light quarks within the heavy hadrons.}
The number of independent contact-range couplings depends on the
ways to combine the light spins $S_{L1}$ and $S_{L2}$ of
the two heavy hadrons $1$ and $2$:
$S_{L1} \otimes S_{L2} = |S_{L1} - S_{L2} | \oplus \ldots \oplus (S_{L1} + S_{L2})$.
This means, for instance, that in the $D \bar{D}$ and $\Sigma_c \bar{D}$
families of molecules there are two independent couplings,
in the $\Sigma_c \Sigma_c$ family three independent
couplings and in the $D_1 \bar{D}_1$ family four couplings.
In addition, if the two heavy hadrons have different light spin,
there is the possibility of additional couplings from operators
involving the exchange of light spin
(the $\Lambda_c \Sigma_c$ system being an example).
From this the S-wave contact-range interaction of two heavy hadrons
can be written as
\begin{eqnarray}
  V_C = C_0 + C_1\,\hat{S}_{L1} \cdot \hat{S}_{L2} +
  C_2\,\hat{Q}_{L1, ij} \hat{Q}_{L2, ij} + \dots, \label{eq:series}
\end{eqnarray}
that is, a series of the products of irreducible tensors built
from the light{-}spin operators $\vec{S}_{L1}$ and $\vec{S}_{L2}$.
The operator $\hat{S}_{L} = \vec{S}_{L} / | \vec{S}_L |$
is a normalized spin operator,
while the operator $\hat{Q}_{L, ij}$ is the spin-2 product
\begin{eqnarray}
  Q_{L,ij} = \frac{1}{2}\left[ S_{L,i} S_{L,j} + S_{L, j} S_{L, i} \right]
  - \frac{\vec{S}_L^2}{3} \delta_{i,j} \, ,
\end{eqnarray}
which is later normalized as $\hat{Q}_{L,ij} = Q_{L, ij} / |Q_{L, 33}|$.
Analogously we can define higher{-}spin products of $S_{L1}$ and $S_{L2}$.

To determine how to saturate the couplings $C_J$ of the effective potential,
we will {split} it in two contributions coming from the scalar- and
vector-meson potentials: $C_J = C_J^S + C^V_J$.
We begin by {writing} the Lagrangians.
For the interaction of a scalar meson with the light{-}quark degrees of freedom,
the Lagrangian reads
\begin{eqnarray}
  \mathcal{L}^S = g_{\sigma} \, {q}_L^{\dagger} \sigma \, q_L \, ,
\end{eqnarray}
where $g_{\sigma}$ is a coupling constant, $\sigma$ is the scalar meson field
\mpv{ and $q_L$ is a non-relativistic field with the quantum numbers of
the light quarks within the heavy hadron, i.e. instead of
writing down the full heavy-hadron field, what we are using is
an effective field that only contains the degrees of freedom
that are relevant for describing interactions
among heavy hadrons.}
With this Lagrangian we end up with the potential
\begin{eqnarray}
  V_{\sigma} = - \frac{g_{\sigma}^2}{q^2 + m_{\sigma}^2} \, ,
\end{eqnarray}
for which saturation reads
\begin{eqnarray}
  C^S_0 \sim - \frac{g_{\sigma}^2}{m_{\sigma}^2} \, , \quad C^S_{J \geqslant 1} \sim 0 \, .
\end{eqnarray}
For the vector mesons the Lagrangian can be written as the multipole expansion
\begin{eqnarray}
  \mathcal{L}^V =&& \,\mathcal{L}^V_{E0} + \mathcal{L}^V_{M1} + \mathcal{L}^V_{E2}
  + \dots \nonumber \\
  =&& \,g_{V} \, {q}_L^{\dagger} V_0 q_L  
  + \frac{f_{V}}{2 M} \, q_L^{\dagger} \,
\hat{S}_{L}\cdot (\vec{\nabla}\times\vec{V})\, q_L
  \nonumber \\
  &&+\, \frac{h_{V}}{2 M^2} \, {q_L}^\dagger \,
  \hat{Q}_{L{\color{blue},}i j} \, \partial_i \partial_{j} V_0 \, q_L + \dots \, ,
\end{eqnarray}
where we have explicitly written the electric charge, magnetic dipole and
electric quadrupolar terms and with the dots indicating higher{-}order
multipole terms.
In this Lagrangian, $g_V$, $f_V$ and $h_V$ are coupling constants,
$V_{\mu} = (V_0, \vec{V}\,)$ is the vector meson field and $M$ is
the typical mass scale associated to the size of
the vector mesons.
The number of terms depends on the spin of the light{-}quark degrees of
freedom, where for $S_L = 0$ (e.g. $\Lambda_c$) there is only the electric term,
for $S_L = \frac{1}{2}$  (e.g. $D$, $D^*$)
there is also the magnetic dipole term,
for $S_L = 1$ ($\Sigma_c$, $\Sigma_c^*$) we add
the electric quadrupole term, and so on.
From this Lagrangian it is easy to derive the {one-boson-exchange}
potential~\cite{Machleidt:1987hj} for a particular
two{-}heavy-hadron system, where the contributions read
\begin{eqnarray}
  V_{E0} &=& +\frac{g_V^2}{q^2 + m_V^2} \, , \\
  V_{M1} &=& -\frac{f_V^2}{4 M^2}\,
  \frac{(\hat{S}_{L1} \times \vec{q}\,) \cdot (\hat{S}_{L2} \times \vec{q}\,)}{q^2 + m_V^2}
  \nonumber \\
  &=& -\frac{2}{3}\,\frac{f_V^2}{4 M^2}\,\hat{S}_{L1} \cdot \hat{S}_{L2}\,
  \frac{q^2}{q^2 + m_V^2} + \dots \, , \\
  V_{E2} &=& +\frac{h_V^2}{4 M^4}\, 
  \frac{(\hat{Q}_{L1,ij}\,q_i q_j) \,
    (\hat{Q}_{L2, lm}\,q_l q_m)}{q^2 + m_V^2} \nonumber \\
  &=& +\frac{h_V^2}{36 M^4}\,(\hat{Q}_{L1, ij} \hat{Q}_{L2, ij})\,
  \frac{q^4}{q^2 + m_V^2} + \dots \, , 
\end{eqnarray}
where for the M1 and E2 terms we isolate the S-wave piece in the second line.
If we remove the Dirac-delta terms, we can deduce the saturation condition
for vector{-}meson exchange.
But first we have to distinguish between the $\omega$ and $\rho$ meson
contributions.
The most obvious difference is that the $\rho$ contribution contains
an isospin factor that we have not explicitly written.
Owing to the negative G-parity of the $\omega$, its contribution changes
sign depending on whether we are dealing with a hadron-hadron or
hadron-antihadron system.
Regarding the couplings, SU(3)-flavor symmetry and the OZI rule
imply that the $\rho$ and $\omega$ couplings are identical
for heavy{ }hadrons in the $3$ or $6$ representation
(which include all the cases considered here).
After removing the Dirac-delta terms, we get the saturation conditions
\begin{eqnarray}
  C_0^V(\mu \sim m_V) &\sim& \frac{g_V^2}{m_V^2}\,(
    \zeta + \bm{\hat{T}}_1 \cdot \bm{\hat{T}}_2 ) \, , \\
  C_1^V(\mu \sim m_V) &\sim& \frac{f_V^2}{6 M^2}\,(
      \zeta + \bm{\hat{T}}_1 \cdot \bm{\hat{T}}_2) \, , \\
  C_2^V(\mu \sim m_V) &\sim& \frac{h_V^2 m_V^2}{36 M^4}\,(
    \zeta + \bm{\hat{T}}_1 \cdot \bm{\hat{T}}_2 ) \, ,
\end{eqnarray}
where $\zeta = \pm 1$ gives the contribution from the omega
and $\bm{\hat{T}}_i = \bm{T}_i / T_i$ is the normalized isospin operator.
The saturation condition generates $C_J$ couplings with consistent signs.
From this we can see that for the isoscalar hadron-antihadron system
the saturated couplings are always attractive:
\begin{eqnarray}
  C_J^{V}(I < I_1 + I_2) < 0 \, .
\end{eqnarray}
This does not imply that the potential is always attractive, because
that will depend on the linear combination of $C_J$'s that conform
the contact-range potential in a given channel.
Yet, if we notice that the $C_J$'s follow a multipole expansion,
the natural expectation is that terms involving higher multipoles
will be smaller:
\begin{eqnarray}
   |C_{J'}^{V}| < |C_{J}^{V}| \quad \mbox{for $J' > J$.}
\end{eqnarray}
This expectation is indeed confirmed by the LHCb pentaquark trio,
provided they are molecular, as attested by a few theoretical
works~\cite{Liu:2018zzu,Liu:2019tjn,Valderrama:2019chc,Du:2019pij}.

\begin{table}[!ttt]
\begin{tabular}{|cccc|}
\hline\hline
  Molecule  & $J^{P}$ & $V$ & Attractive? \\
  \hline
  $D \bar{D}$ & $0^{++}$ & $C_0$ & Yes \\ \hline
  $D^* \bar{D}$ & $1^{++}$ & $C_0 + C_1$ & Most \\ 
  $D^* \bar{D}$ & $1^{+-}$ & $C_0 - C_1$ & Likely \\ \hline
  $D^* \bar{D}^*$ & $0^{++}$ & $C_0 -2 C_1$ & Likely \\
  $D^* \bar{D}^*$ & $1^{+-}$ & $C_0 - C_1$ & Likely \\ 
  $D^* \bar{D}^*$ & $2^{++}$ & $C_0 + C_1$ & Most \\ \hline
  \hline
  Molecule  & $J^{P}$ & $V$ & Attractive? \\
  \hline
  $\bar{D} \Sigma_c$ & $\frac{1}{2}^-$ & $D_0$ & Yes \\ \hline
  $\bar{D} \Sigma_c^*$ & $\frac{3}{2}^-$ & $D_0$ & Yes\\ \hline
  $\bar{D}^* \Sigma_c$ & $\frac{1}{2}^-$ & $D_0 - \frac{4}{3}\,D_1$ & Likely \\
  $\bar{D}^* \Sigma_c$ & $\frac{3}{2}^-$ & $D_0 + \frac{2}{3}\,D_1$ & Most \\
  \hline
  $\bar{D}^* \Sigma_c^*$ & $\frac{1}{2}^-$ & $D_0 - \frac{5}{3}\,D_1$
  & Likely \\
  $\bar{D}^* \Sigma_c^*$ & $\frac{3}{2}^-$ & $D_0 - \frac{2}{3}\,D_1$
  & Likely \\
  $\bar{D}^* \Sigma_c^*$ & $\frac{5}{2}^-$ & $D_0 + D_1$ & Most \\
  \hline \hline
  Molecule  & $J^{P}$ & $V$ & Attractive? \\
    \hline
  $\Sigma_c \bar{\Sigma}_c$ & $0^{-+}$ & $E_0 - \frac{4}{3} E_1$ & Likely \\ 
  $\Sigma_c \bar{\Sigma}_c$ & $1^{--}$ & $E_0 + \frac{4}{9} E_1$ & Yes \\ \hline
    $\Sigma_c^* \bar{\Sigma}_c$ & $1^{-+}$ & $E_0 - E_1 - \frac{15}{2} E_2$ &
    Likely \\
  $\Sigma_c^* \bar{\Sigma}_c$ & $1^{--}$ &
  $E_0 - \frac{11}{9} E_1 + \frac{15}{2} E_2$ & Likely \\
  $\Sigma_c^* \bar{\Sigma}_c$ & $2^{-+}$ &
  $E_0 + \frac{1}{3} E_1 - \frac{3}{2} E_2$ & Likely \\
  $\Sigma_c^* \bar{\Sigma}_c$ & $2^{--}$ &
  $E_0 + E_1 + \frac{3}{2} E_2$ & Most \\ \hline
  $\Sigma_c^* \bar{\Sigma}_c^*$ & $0^{-+}$ &
  $E_0 - \frac{15}{9} E_1 + \frac{15}{2} E_2$ & Likely \\
  $\Sigma_c^* \bar{\Sigma}_c^*$ & $1^{--}$ &
  $E_0 - \frac{11}{9} E_1 + \frac{3}{2} E_2$ & Likely \\
  $\Sigma_c^* \bar{\Sigma}_c^*$ & $2^{-+}$ &
  $E_0 - \frac{1}{3} E_1 - \frac{9}{2} E_2$ & Likely \\
  $\Sigma_c^* \bar{\Sigma}_c^*$ & $3^{--}$ &
  $E_0 + E_1 + \frac{3}{2} E_2$ & Most \\ \hline 
  \hline\hline 
\end{tabular}
\caption{Structure of the contact-range potential for the $\bar{D} D$,
  $\bar{D} \Sigma_c$ and $\bar{\Sigma}_c \Sigma_c$ family of
  molecules. For configurations in which the isospin $I$ of the molecule
  is not maximal, $I < I_1 + I_2$, all the couplings appearing
  in this table are expected to be negative in sign (i.e. attractive).
  If we take into account that the previous couplings are expected
  to be smaller as the multipole moment increases, then we arrive
  at the labels ``Most'', ``Yes'', ``Likely'' to characterize
  whether a particular molecular configuration is attractive.
}
\label{tab:molecules}
\end{table}

To illustrate this idea we consider a few examples:
(1) the $D D$ and $D \bar{D}$ family of molecules,
(2) the $D \Sigma_c$ and $\bar{D} \Sigma_c$ family, and
(3) the $\Sigma_c \Sigma_c$ and $\Sigma_c \bar{\Sigma}_c$ one.
We have summarized the form of the contact-range potential
for these three cases in Table \ref{tab:molecules}.
For the first case, which includes the $X(3872)$,
it is more convenient to define the contact-range
potential in terms of the Pauli matrices (instead of the spin matrices)
\begin{eqnarray}
  V_c^{(1)} = C_0 + \vec{\sigma}_{L1} \cdot \vec{\sigma}_{L2} \, C_1 \, ,
\end{eqnarray}
for which vector saturation gives
\begin{eqnarray}
  C_0^V &\sim&
  \frac{g_V^2}{m_V^2} (\zeta + \bm{\tau}_1 \cdot \bm{\tau}_2 ) \, , 
\end{eqnarray}
plus the analogous expression for $C_1^V$.
From this it is clear that the $I=0$ isoscalar configurations
are guaranteed to be attractive.
For the isovector configurations the $\rho$ and $\omega$ contributions
cancel out: for the $C_0$ coupling there is still the scalar-meson
contribution, which will result in attraction, while for the $C_1$
coupling the sign will depend on how the SU(3)-flavor symmetry
is broken.
Alternatively, the exchange of the $a_1$ meson~\cite{Durso:1984um} would imply
$C_1(I_1 = 1) \geqslant 0$ for the $Z_c(3900)$ and $Z_c(4020)$ resonances,
which is compatible with their quantum numbers ($J^{PC} = 1^{+-}$).
Thus it might be possible that the $I = I_1 + I_2$ configurations
revert to the naive expectation of higher (light-quark) spin
states having higher masses.

For the second case, the $\bar{D} \Sigma_c$ and $D \Sigma_c$ family of molecules
(which include the LHCb pentaquark trio), we define the contact-range
potential as
\begin{eqnarray}
  V_C^{(2)} = D_0 + \vec{\sigma}_{L1} \cdot \vec{S}_{L2} \, D_1 \, ,
\end{eqnarray}
where $\vec{S}_{L2}$ refers to the spin-1 angular momentum matrices.
Saturation in this case gives
\begin{eqnarray}
  D_0^V &\sim&
  \frac{g_V g_V'}{m_V^2} (\zeta + \bm{\tau}_1 \cdot \bm{T}_2 ) \, , 
\end{eqnarray}
plus the analogous expression for $D_1$, with $g_V'$ the vector-meson
coupling for the $\Sigma_c$ and $\Sigma_c^*$ baryons and {$\bm{T}_2$}
their isospin operators.
This expression indicates that the isospin-$\frac{1}{2}$ configurations are
attractive for both the $\bar{D} \Sigma_c$ and $D \Sigma_c$ cases.
A second conclusion is that in the $\bar{D}^* \Sigma_c$ system
the $J^P = \tfrac{3}{2}^-$ configuration is expected
to be more attractive than the $J^P = \tfrac{1}{2}^-$ one, which implies
that the quantum numbers of the $P_c(4440)$ and $P_c(4457)$ pentaquarks
should be $J^P = \tfrac{3}{2}^-$ and $\tfrac{1}{2}^-$, respectively.
A third conclusion is that the doubly charmed ${D} \Sigma_c$-type
family of molecules are expected to be more tightly bound
than the hidden{-}charm pentaquarks, owing to the different
sign of the $\omega$ contribution~\cite{Yu:2019yfr}.

Finally, if we apply it to the $\Sigma_c \Sigma_c$ and
$\Sigma_c \bar{\Sigma}_c$ family of molecules,
the contact-range potential reads
\begin{eqnarray}
  V_C^{(3)} =
  E_0 + E_1\,\vec{S}_{L1} \cdot \vec{S}_{L2} +
  E_2\,\hat{Q}_{L1,ij} \hat{Q}_{L2,ij} \, .
\end{eqnarray}
The vector{-}meson saturation of the couplings yields
\begin{eqnarray}
  E_0 &\sim&
  \frac{g_V'^2}{m_V^2} (\zeta + \bm{T}_1 \cdot \bm{T}_2 ) \, ,
\end{eqnarray}
plus the analogous expressions for $E_1$ and $E_2$.
From this the isoscalar and isovector $2^{--}$ and $3^{--}$
heavy baryonia are expected to be the most attractive.

\mpv{
  We stress the qualitative character of the present analysis.
  Saturation requires two conditions for the regularization scale $\mu$:
  it must be close to the mass of the exchanged meson $m$ and
  it must be (ideally much) softer than the physical cutoff $M$,
  i.e. $\mu \sim m$ and $\mu < M$ (even better: $\mu \ll M$).
  Though these two conditions are indeed met for scalar- and vector-meson
  exchange, the ratio $\mu / M$ is not small, which indicates
  that the saturation of the couplings is not necessarily
  expected to do well quantitatively.
  However previous investigations on the couplings
  in the pion-nucleon~\cite{Ecker:1988te} and
  nucleon-nucleon~\cite{Epelbaum:2001fm} systems indicate
  that contact-range couplings are indeed saturated
  by light-meson exchange.
  We do not know whether this will be the case for hadronic molecules, yet
  the present manuscript focuses on the qualitative aspects of saturation,
  particularly the signs of the couplings, which are more likely
  to be unaffected by the poor scale separation.
}

To summarize, we propose a description of heavy-hadron molecules
in terms of contact-range potentials that depend on a few couplings.
The couplings are determined from saturation by scalar{-} and
vector{-}meson exchange, where we propose a novel saturation
procedure that takes into account the physical scale
at which saturation is actually happening.
The outcome is that it is possible to know the sign and relative strength of
the two{-}heavy-hadron interaction, from which we can deduce a few qualitative
properties of the heavy molecular spectrum.
The most interesting pattern is that for heavy molecular states without
maximal isospin, we expect the configurations
with higher light-quark spin to be more bound
(or, equivalently, lighter if we {refer 
to} the mass of the states).
This pattern is exactly the opposite {of the} one that is observed
in standard compact hadrons, where mass usually increases with spin.
The pattern is compatible with the quantum numbers of the $X(3872)$
in the molecular picture and with the experimental absence of
molecular partners of the $X(3872)$
with smaller light-quark spin.
The pattern also extends to the light sector, with the deuteron
(neutron-proton, $I(J) = 0(1)$) and the recently observed
$d^*(2380)$ ($\Delta \Delta$, $I(J)=0(3)$)
being two illustrative examples.
Yet the real test of the present idea will be the eventual experimental
measurement of the quantum numbers of the $P_c(4440)$ and $P_c(4457)$
pentaquarks.
If they are $\bar{D}^* \Sigma_c$ molecules, the saturation hypothesis suggests
that the $J = \tfrac{3}{2}$ state should be the most bound of the two,
i.e. the spin of the $P_c(4440)$ should be $\tfrac{3}{2}$.

\section*{Acknowledgments}

This work is partly supported by the National Natural Science Foundation
of China under Grants No. 11735003, 11975041, the Thousand Talents Plan
for Young Professionals and the Fundamental Research Funds
for the Central Universities.

\appendix
\section{Scattering length and saturation in a contact-range theory}
\label{app:pionless}

Here we explain the calculation of the scattering length $a_0$ and
the choice of the regularization scale $\mu$ that
we have presented in Fig.~\ref{fig:saturation}.
First we explicitly regularize the contact-range potential of Eq. \eqref{eq:VC}, i.e.
\begin{eqnarray}
  \braket{\vec{p}{\,'}|V^R_C|\vec{p}\,} &=&
  f_R(p'/\mu)\,\left[ C_0(\mu) + C_1(\mu)\,\vec{\sigma}_1 \cdot \vec{\sigma}_2
  \right]
  \,f_R(p/\mu) \, , \nonumber \\
\end{eqnarray}
where we have chosen a generic non-local regulator $f_R(x)$ such that
$f_R(0) = 1$ and $f_R(x \to \infty) \to 0$.
For obtaining the scattering matrix $T$, we insert the regularized
contact-range potential $V_C^R$ in the Lippmann-Schwinger equation
\begin{eqnarray}
  T(E_{\rm cm}) = V_C^R + V_C^R \, G_0(E_{\rm cm}) \, T(E_{\rm cm}) \, ,
\end{eqnarray}
with $E_{\rm cm}$ the center-of-mass energy of the two-body system and
$G_0(E_{\rm cm}) = 1 / (E_{\rm cm} - H_0)$ the resolvent operator,
$H_0$ being the free Hamiltonian (i.e. the kinetic energy operator).
As we are interested in the scattering length, we simply take 
\begin{eqnarray}
  T(E_{\rm cm} = 0) = \frac{2 \pi \, a_0}{\mu_H} \, , 
\end{eqnarray}
with $\mu_H$ the reduced mass of the two-hadron system and
$a_0$ the scattering length.
In this limit the Lippmann-Schwinger equation simplifies to
\begin{eqnarray}
  \frac{\mu_H}{2 \pi \, a_0} = {[C_0(\mu)+ C_1(\mu)\,\vec{\sigma}_1 \cdot \vec{\sigma}_2]}^{-1} - I_0(0, \mu) \, .
\end{eqnarray}
Here $I_0(k, \mu)$ is the loop integral
\begin{eqnarray}
I_0(k,\mu) &=& \int\frac{\mathrm{d}^3q}{{(2\pi)^3}}\, \frac{f_R^2(q/\mu)}{(k^2-q^2)/{2\mu_H}+i0^+}\notag\\
&=& -\frac{\mu_H}{2 \pi}\left[ ik + \beta\mu + \mathcal{O}(k^2/\mu)\right]\,,
\end{eqnarray}
with $k$ the center-of-mass momentum ($k = \sqrt{2 \mu_H E_{cm }}$) and
$\beta=\mathcal{O}(1)$ a regulator-dependent number,
e.g. for a sharp-cutoff (Gaussian)
regulator $f_R(x)=\theta(1-x)$ ($f_R(x) = e^{-x^2/2}$)
we end up with $\beta=2/\pi$ ($\beta=1/\!\sqrt{\pi}$).

We will consider the case in which the underlying theory to which
we want to match $C_0$ and $C_1$ is a Yukawa potential of the type
\begin{eqnarray}
  V_Y(\vec{q}\,) = - \lambda\,\frac{m_Y^2}
  {\vec{q}^{\,2} + m_Y^2} \, ,
\end{eqnarray}
where $m_Y$ is the mass of the exchanged boson and $\lambda$ a coupling constant
with dimensions of $[{\rm mass}]^{-2}$.
If the following condition is met~\cite{SanchezSanchez:2017xtl}
\begin{eqnarray}
  \frac{\mu_H}{2 \pi} \, m_Y \lambda \simeq 1.68 \, , \label{eq:Y-th}
\end{eqnarray}
then the Yukawa potential will have a bound state at threshold.
Additionally we will write the coupling $\lambda$ as
\begin{eqnarray}
  \lambda = -\tau \frac{g^2}{6 f_{\pi}^2} \,
  \vec{\sigma}_1 \cdot \vec{\sigma}_2 \, , \label{eq:lambda}
\end{eqnarray}
by which we reproduce the calculation
of Fig.~\ref{fig:saturation}.

Actually the coupling $\lambda$ for which the Yukawa potential has a bound
state at threshold (i.e. Eq.~(\ref{eq:Y-th})) provides a good matching
point for saturating the contact-range couplings $C_0$ and $C_1$.
A bound state at threshold is equivalent to the limit in which
the scattering length diverges, $a_0 \to \infty$,
for which the couplings should be
\begin{eqnarray}
  C_0(\mu)+ C_1(\mu)\,\vec{\sigma}_1 \cdot \vec{\sigma}_2 = -
  \frac{2 \pi}{\mu_H}\,\frac{1}{\beta \mu} \, .
\end{eqnarray}
If we impose saturation of the couplings
\begin{eqnarray}
  C_0(\mu)+ C_1(\mu)\,\vec{\sigma}_1 \cdot \vec{\sigma}_2 = -\lambda \, ,
\end{eqnarray}
which given Eq.~(\ref{eq:lambda}) is equivalent to
\begin{eqnarray}
  C_0(\mu) = 0 \quad \mbox{and} \quad C_1(\mu) = \tau \frac{g^2}{6 f_{\pi}^2}
  \, ,
\end{eqnarray}
we end up with the following condition for the regularization scale $\mu$
at which exact saturation happens for a Yukawa potential
\begin{eqnarray}
  \mu_{\rm sat} \simeq \frac{m_Y}{1.68\, \beta} \, .
\end{eqnarray}
If we particularize this condition for a sharp-cutoff (Gaussian) regulator,
we get $\mu_{\rm sat} \simeq 0.93\,m_Y$ ($\mu_{\rm sat} \simeq 1.06\,m_Y$),
which satisfies our original expectation that saturation
works for $\mu \sim m_Y$.
For the example we give in the main text, the OPE potential, the saturation
scale will be $\mu_{\rm sat} \simeq 131\,{\rm MeV}$ for the sharp-cutoff case,
thus reproducing the scale at which Fig.~\ref{fig:saturation} is calculated
(while for a Gaussian regulator we would have obtained
$\mu_{\rm sat} \simeq 148\,{\rm MeV}$ instead).
For the exchange of heavier light mesons ($\sigma$, $\rho$, $\omega$)
the potential will not be Yukawa-like owing to finite hadron size
effects, which cannot be ignored in this latter case.
Thus the type of clean saturation relations we have derived here
should only be expected to be valid at the qualitative level.

\section{Heavy-superfield and light-subfield notations}
\label{app:H}

In this Appendix we explain the non-standard notation
we use for heavy hadrons throughout this manuscript.
Heavy hadrons are composed of heavy and light quarks ($Q$ and $q$),
but from HQSS we expect that their properties and interactions
will be independent of the combined spin of the heavy quarks.
If the spin of the heavy- and light-quarks within a heavy hadron
is $S_H$ and $S_L$ respectively, the spin of the heavy-hadron
can be $S = |S_H - S_L| \oplus\, \ldots\, \oplus (S_H + S_L)$.
The only difference between these combination is how the heavy- and light-quark
spins couple, but the properties of the resulting heavy hadron will only
depend on $S_L$.
The standard way to take this into account is to combine the different
heavy hadrons with the same light-quark spin into multiplets
with good properties with respect to rotations of $S_H$.
For example, if we are considering the charmed mesons $D$, $D^*$
with total spin $S = 0$, $1$ respectively, and heavy- and light-quark
spins $S_H = 1/2$ and $S_L = 1/2$, it is customary to group them
into the superfield
\begin{eqnarray}
  H_c = \frac{1}{\sqrt{2}}\left[
    {\bf 1}\,D + \vec{\sigma} \cdot \vec{D}^* \right] \, ,
\end{eqnarray}
with ${\bf 1}$ and $\vec{\sigma}$ the 2{$\times$}2 identity and Pauli matrices
respectively, where this specific representation corresponds to
the non-relativistic limit of the one used
in Ref.~\cite{Falk:1992cx}.
The superfield $H_c$ transforms as $H_c \to e^{- i \vec{S}_H \cdot \vec{\theta}} H_c$
under a rotation $\vec{\theta}$ of the heavy-quark spin $\vec{S}_H$,
while this rotation mixes the $D$ and $D^*$ fields. 

Now if we want to construct a Lagrangian for contact-range interactions without
derivatives for the charmed mesons, we just have to write this Lagrangian
in terms of the superfield $H_c$ to ensure HQSS, where the result is
\begin{eqnarray}
  \mathcal{L}_{4 H} &=&
  C_a {\rm Tr}\left[ H_c^{\dagger} H_c \right] {\rm Tr}\left[ H_c^{\dagger} H_c \right] \nonumber \\ &+&
  C_b \sum_i {\rm Tr}\left[ H_c^{\dagger} \sigma_i H_c \right] {\rm Tr}\left[ H_c^{\dagger} \sigma_i H_c \right]  \, , \label{eq:L-4H}
\end{eqnarray}
{\mpv``Tr'' standing for the trace computed over the spin indices.}
Expanding the superfield $H_c$ in terms of the charmed-meson fields $D$ and $D^*$, we will obtain the Lagrangian
\begin{eqnarray}
  \mathcal{L}_{4 H} &=&
  C_a {\left( D^{\dagger} D + \vec{D}^{*\dagger} \cdot \vec{D} \right)}
  {\left( D^{\dagger} D + \vec{D}^{*\dagger} \cdot \vec{D} \right)}
  \nonumber \\
  &+& C_b {\left( D^{\dagger} \vec{D}^{*} + \vec{D}^{*\dagger} {D} \right)}
  \cdot {\left( D^{\dagger} \vec{D}^{*} + \vec{D}^{*\dagger} {D} \right)}
  \nonumber \\
  &-& i \, C_b \Big[
    {\left( D^{\dagger} \vec{D}^{*} + \vec{D}^{*\dagger} {D} \right)}
  \cdot {\left( \vec{D}^{*\dagger} \times \vec{D}^{*} \right)}
  \nonumber \\
  && \phantom{i C_b} - {\left( \vec{D}^{*\dagger} \times \vec{D}^{*} \right)}
  \cdot
  {\left( D^{\dagger} \vec{D}^{*} + \vec{D}^{*\dagger} {D} \right)} \Big]
    \nonumber \\
  &-& C_b {\left( \vec{D}^{*\dagger} \times \vec{D}^{*} \right)}
  \cdot {\left( \vec{D}^{*\dagger} \times \vec{D}^{*} \right)} \, ,
\end{eqnarray}
from which we can deduce the potentials
for the $DD$, $DD^*$ and $D^*D^*$ cases
\begin{eqnarray}
  V_C(DD) &=& C_a \, , \label{eq:V-DD} \\
  V_C(DD^* \pm D^* D) &=&
  C_a  \pm C_b \, {\vec{\epsilon}\,}^* \cdot \vec{\epsilon} \, , \\
  V_C(D^* D^*) &=& C_a  + C_b \, \vec{S}_1 \cdot \vec{S}_2 \, ,
  \label{eq:V-DbarDbar}
\end{eqnarray}
where $\vec{\epsilon}$ refers to the polarization vector of the $D^*$ charmed
meson, $\vec{S}_i$ to the spin-1 matrices for the $i=1,2$ meson,
and the sign of the $DD^*$ potential depends on whether we have
a symmetric or antisymmetric combination of the two mesons.

Alternatively, if we notice that the heavy-quark spin degrees of freedom
do not appear in the interaction between two heavy hadrons,
then we can simplify the derivation of the potential.
The point is that instead of grouping the $D$ and $D^*$ fields into
the superfield $H_c$, we can simply strip down the heavy-quark spin
from the $D$ and $D^*$ fields to write a simplified subfield
only containing the light-quark spin degrees of freedom
\begin{eqnarray}
  D \, , \, D^* \quad \longrightarrow \quad q_L \, , \, \vec{\sigma}_L \, ,
\end{eqnarray}
where $q_L$ represents the subfield and $\vec{\sigma}_L$ is
the spin of the light-quark within the $D$ and $D^*$.
With this notation the contact-range Lagrangian of Eq.~(\ref{eq:L-4H})
now reads
\begin{eqnarray}
  \mathcal{L}_{4 H}
  &=& C_a \, (q_L^{\dagger} \, q_L) \, (q_L^{\dagger} \, q_L)
  \nonumber \\
  &+& C_b \, \sum_i (q_L^{\dagger} \sigma_{Li} q_L) \,
  (q_L^{\dagger} \sigma_{Li} q_L) \, ,
\end{eqnarray}
from which we can directly obtain the potential
\begin{eqnarray}
  V_C = C_a + C_b\,\vec{\sigma}_{L1} \cdot \vec{\sigma}_{L2} \, .
\end{eqnarray}
The only difficulty are the matrix elements of the operator $\vec{\sigma}_{L}$
when sandwhiched between the charmed-meson fields, but these
can be readily obtained from the coupling of the heavy meson and
heavy- and light-quark spins, yielding
\begin{eqnarray}
  \langle D | \vec{\sigma}_L | D \rangle &=& 0 \, , \\
  \langle D | \vec{\sigma}_L | D^* \rangle &=& \vec{\epsilon} \, , \\
  \langle D^* | \vec{\sigma}_L | D^* \rangle &=& \vec{S} \, ,
\end{eqnarray}
from which we reproduce the potentials of
Eqs.~(\ref{eq:V-DD}-\ref{eq:V-DbarDbar}).

%

\end{document}